\documentclass[preprintnumbers, floatfix, letterpaper, twocolumn,aps,prd,epsfig,nofootinbib,natbib,longbibliography]{revtex4-1}
\usepackage{bm,graphicx,dcolumn,epstopdf,epsf, latexsym,mathbbol, amssymb,amsmath,color,slashed, mathrsfs,mathcomp,simplewick}
\usepackage[center]{subfigure}
\usepackage{multirow}
\usepackage{makecell}
\usepackage[colorlinks,linkcolor=blue,citecolor=blue,urlcolor=blue]{hyperref}

\begin{document}
\allowdisplaybreaks
 \newcommand{\bq}{\begin{equation}}
 \newcommand{\eq}{\end{equation}}
 \newcommand{\bqn}{\begin{eqnarray}}
 \newcommand{\eqn}{\end{eqnarray}}
 \newcommand{\nb}{\nonumber}
 \newcommand{\lb}{\label}
\newcommand{\f}{\frac}
\newcommand{\p}{\partial}
\newcommand{\PRL}{Phys. Rev. Lett.}
\newcommand{\PLB}{Phys. Lett. B}
\newcommand{\PRD}{Phys. Rev. D}
\newcommand{\CQG}{Class. Quantum Grav.}
\newcommand{\JCAP}{J. Cosmol. Astropart. Phys.}
\newcommand{\JHEP}{J. High. Energy. Phys.}

\title{Gravitational plane waves  in Einstein-aether theory}

 \author{Jacob Oost$^{a}$}
\email{Jacob$\_$Oost@baylor.edu}

 \author{Madhurima Bhattacharjee$^{a}$}
\email{Madhurima$\_$Bhattacharjee@baylor.edu}

 \author{Anzhong Wang$^{a, b}$}
\email{Anzhong$\_$Wang@baylor.edu; Corresponding Author}

\affiliation{$^{a}$ GCAP-CASPER, Physics Department, Baylor
University, Waco, TX 76798-7316, USA  \\
$^{b}$ Institute  for Advanced Physics $\&$ Mathematics,   Zhejiang University of
Technology, Hangzhou 310032,  China }

\date{\today}

\begin{abstract}
 
In this paper,   we systematically study spacetimes of gravitational plane waves in Einstein-aether theory. Due to the presence of 
the timelike aether vector field,  now the problem in general becomes overdetermined. In particular, for the linearly polarized  plane 
waves, there are  five independent vacuum Einstein-aether  field equations for three unknown functions. Therefore, solutions exist 
only for particular choices of the four free parameters $c_{i}$'s of the theory.  We find that there exist eight cases, in two of which 
any form of gravitational plane waves can exist, similar to that in general relativity, while in the other six cases,  gravitational plane 
waves exist only in particular forms. Beyond these eight cases,  solutions either do not exist or are trivial (simply representing  a 
Minkowski spacetime with a constant or dynamical aether field.).

\end{abstract}


\maketitle

\section{Introduction}
 \renewcommand{\theequation}{1.\arabic{equation}} \setcounter{equation}{0}

Lorentz invariance (LI) has been the cornerstone of modern physics and is strongly supported by observations \cite{Kostelecky:2008ts}. In fact, all the 
experiments carried out so far are consistent with it, and there is no evidence to show that such a symmetry needs to be broken at a certain energy scale, 
although the constraints on the LI violations in the matter sector are much stronger than those in the gravitational sector \cite{M-L}. 

Nevertheless, there are various reasons to construct gravitational theories with broken LI. In particular, when spacetime is quantized, as what we currently 
understand from the point of view of quantum gravity \cite{Carlip,Kiefer}, space and time emerge from some discrete substratum.  Then,  LI, as a continuous 
spacetime symmetry, cannot exist in such discrete space and time. Therefore,  it cannot be a fundamental symmetry,  instead  must be an emergent one 
at the low energy physics. Following this line of thinking, various gravitational theories that violate LI have been proposed, such as ghost condensation 
\cite{ArkaniHamed:2003uy}, Einstein-aether theory \cite{Jacobson}, and more recently, Ho\v{r}ava  theory of gravity \cite{Horava}. 
 While the ghost condensation and Einstein-aether theory are considered as the low energy effective theories of gravity, the Ho\v{r}ava gravity is supposed 
 to be ultraviolet (UV) complete \cite{Wang17}. 
 
 In Einstein-aether theory,   LI is broken only down to a rotation subgroup by the existence of a preferred time direction at every point of spacetime, i.e., the existence of a 
 preferred frame of reference established by the aether vector field. This time-like unit vector field  can be interpreted as a velocity four-vector of some medium substratum 
 (aether, vacuum, or dark fluid), bringing into consideration non-uniformly-moving continuous media and their interaction with other fields. Meanwhile, this theory can be
also considered as  a realization of dynamic self-interaction of complex systems moving with a spacetime dependent macroscopic velocity.

The introduction of the aether vector allows for some novel effects, e.g., matter fields can travel faster than the speed of light \cite{jacobson3}, and new gravitational wave 
polarizations can spread at different speeds \cite{jacobson4}. It should be noted  that  the faster-than-light propagation does not violate causality \cite{Wang17}. 
In particular, gravitational theories with broken LI still allow the existence of black holes \cite{UHs}. However, instead of Killing horizons, 
now the boundaries of black holes are  hypersurfaces termed {\em universal horizons}, which can trap excitations traveling at arbitrarily high velocities (For more details,
see, for example, \cite{Wang17} for  a recent review.).
This universal horizon may radiate thermally at a fixed temperature and strengthen a possible thermodynamic interpretation though there is no universal light cone \cite{berglund2}.

Another interesting  issue is whether or not spacetimes of gravitational plane waves are compatible with the presence of the timelike aether field.  This  becomes more 
interesting  after the recent observations of several gravitational waves (GWs) emitted from remote binary systems of either black holes \cite{GW150914,GW151226,GW170104,GW170814} 
or neutron stars \cite{GW170817}. The sources of these GWs are far from us, and when they arrive to us,  they can be well approximated by gravitational plane waves. However, 
this issue  is not trivial, specially for the Einstein-aether theory, in which a globally time-like aether field exists, while such plane waves, by definition, move along congruences
defined by a null vector. 

In this paper, we shall focus ourselves on this issue. In particular, we shall show that the system of the differential equations for gravitational plane waves in the Einstein-aether theory is
in general overdetermined, that is, we have more independent differential equations than the number of independent functions that describe the spacetime and aether, sharply
in contrast to that encountered in Einstein's General Relativity (GR), in which the problem is usually underdetermined, that is, we have less independent differential equations than the number 
of independent functions that describe the spacetime  \cite{ Wang91th,SKMHH,Griffiths16}. In particular, for the linearly polarized gravitational plane waves, there are five independent vacuum field
equations for three unknown functions  in the Einstein-aether theory, while there is only one independent vacuum field
equation for two  unknown functions  in GR.   

The rest of the paper is organized as follows:  In Sec. II, after briefly presenting the Einstein-aether theory, we give a summary on the gravitational plane waves with two independent polarization directions, 
and define the polarization angle with respect to a parallelly transported basis along the path of the propagating gravitational plane wave. Such a description is valid for any metric theory, including GR and 
Einstein-aether theory. In Sec. III, we systematically study the linearly polarized gravitational plane waves in Einstein-aether theory, and find that such gravitational plane wave solutions exist only for particular choices of the free
parameter $c_i$'s of the theory. We identify all these particular cases, and find that there are in total eight cases. Cases beyond these either do not allow such solutions to exist or are trivial, in the sense
that their spacetime is Minkowski (though sometimes with a dynamical aether field). In Sec. IV, we summarize our main results, and present our concluding remarks. There is also an Appendix, in which the
Einstein-aether field equations for  linearly polarized gravitational plane waves are presented.

\section{Einstein-Aether Theory and gravitational plane waves}
 \renewcommand{\theequation}{2.\arabic{equation}} \setcounter{equation}{0}

In this section, we shall give a brief introduction to the Einstein-aether theory \cite{Jacobson}(See also \cite{Wang12} for a brief comment on slowly rotating black holes in Einstein-aether theory and
Ho\v{r}ava theory) and polarizations of gravitational plane waves \cite{Wang91aa,Wang91th}. 
For more details of the Einstein-aether theory, we refer readers to \cite{Jacobson}, while for gravitational plane waves in GR to  \cite{ Wang91th,SKMHH,Griffiths16}. 

\subsection{Einstein-Aether Theory}

In the Einstein-aether theory, the fundamental variables of the gravitational 
sector are \cite{Jacobson}
\bq
\lb{2.0a}
\left(g_{\mu\nu}, u^{\mu}, \lambda\right),
\eq
with the Greek indices $\mu,\nu = 0, 1, 2, 3$, and $g_{\mu\nu}$ is  the four-dimensional metric  of the space-time
with the signature $(-, +,+,+)$ \cite{Foster06,GEJ07,OMW18}. The four-vector  $u^{\mu}$ represents the aether field, and $\lambda$ is a Lagrangian multiplier 
which guarantees that the aether  four-velocity  is always timelike.  The general action of the theory  is given by,
\bq
\lb{2.0}
S = S_{\ae} + S_{m},
\eq
where  $S_{m}$ denotes the action of matter,  and $S_{\ae}$  the gravitational action of the $\ae$-theory, given by
\bqn
\lb{2.1} 
 S_{\ae} &=& \frac{1}{16\pi G }\int{\sqrt{- g} \; d^4x \Big[R(g_{\mu\nu}) + {\cal{L}}_{\ae}\left(g_{\mu\nu}, u^{\alpha}, {\lambda}\right)\Big]},\nb\\  
S_{m} &=& \int{\sqrt{- g} \; d^4x \Big[{\cal{L}}_{m}\left(g_{\mu\nu}, u^{\alpha}; \psi\right)\Big]}.
\eqn
Here $\psi$ collectively denotes the matter fields, $R$    and $g$ are, respectively, the  Ricci scalar and determinant of $g_{\mu\nu}$, 
 and 
\bq
\lb{2.2}
 {\cal{L}}_{\ae}  \equiv - M^{\alpha\beta}_{~~~~\mu\nu}\left(D_{\alpha}u^{\mu}\right) \left(D_{\beta}u^{\nu}\right) + \lambda \left(g_{\alpha\beta} u^{\alpha}u^{\beta} + 1\right),
\eq
 where $D_{\mu}$ denotes the covariant derivative with respect to $g_{\mu\nu}$, and  $M^{\alpha\beta}_{~~~~\mu\nu}$ is defined as
\bqn
\lb{2.3}
M^{\alpha\beta}_{~~~~\mu\nu} = c_1 g^{\alpha\beta} g_{\mu\nu} + c_2 \delta^{\alpha}_{\mu}\delta^{\beta}_{\nu} +  c_3 \delta^{\alpha}_{\nu}\delta^{\beta}_{\mu} - c_4 u^{\alpha}u^{\beta} g_{\mu\nu}.\nb\\
\eqn
Note that here we assume that matter fields couple not only to $g_{\mu\nu}$ but also to the aether field, which in general violates the weak equivalence principle \cite{Jacobson}.     
The four coupling constants $c_i$'s are all dimensionless, and $G $ is related to  the Newtonian constant $G_{N}$ via the relation \cite{CL04},
\bq
\lb{2.3a}
G_{N} = \frac{G }{1 - \frac{1}{2}c_{14}}.
\eq
 
The variations of the total action with respect to $g_{\mu\nu},\; u^{\mu}$ and $\lambda$ yield, respectively, the field equations,
 \bqn
 \lb{2.4a}
 && E^{\mu\nu} = 8\pi G  T^{\mu\nu},\\
 \lb{2.4b}
&&   \AE_{\mu} = 8\pi G  T_{\mu}, \\
   \lb{2.4c}
 &&  g_{\alpha\beta} u^{\alpha}u^{\beta} = -1, 
 \eqn
where
 \bqn
 \lb{2.5}
 E^{\mu\nu} &\equiv& R^{\mu\nu} - \frac{1}{2}g_{\mu\nu}R - T^{\mu\nu}_{\ae},\nb\\
 T^{\mu\nu} &\equiv&  \frac{2}{\sqrt{-g}}\frac{\delta \left(\sqrt{-g} {\cal{L}}_{m}\right)}{\delta g_{\mu\nu}},\nb\\
T_{\mu} &\equiv& - \frac{1}{\sqrt{-g}}\frac{\delta \left(\sqrt{-g} {\cal{L}}_{m}\right)}{\delta u^{\mu}},\nb\\
  T^{\ae}_{\alpha\beta} &\equiv&  
  D_{\mu}\Big[J^{\mu}_{\;\;\;(\alpha}u_{\beta)} + J_{(\alpha\beta)}u^{\mu}-u_{(\beta}J_{\alpha)}^{\;\;\;\mu}\Big]\nb\\
&& + c_1\Big[\left(D_{\alpha}u_{\mu}\right)\left(D_{\beta}u^{\mu}\right) - \left(D_{\mu}u_{\alpha}\right)\left(D^{\mu}u_{\beta}\right)\Big]\nb\\
&& + c_4 a_{\alpha}a_{\beta}    + \lambda  u_{\alpha}u_{\beta} - \frac{1}{2}  g_{\alpha\beta} J^{\delta}_{\;\;\sigma} D_{\delta}u^{\sigma},\nb\\
 \AE_{\mu} & \equiv &
 D_{\alpha} J^{\alpha}_{\;\;\;\mu} + c_4 a_{\alpha} D_{\mu}u^{\alpha} + \lambda u_{\mu}, 
 \eqn
 with
\begin{equation}
 \lb{2.6}
J^{\alpha}_{\;\;\;\mu} \equiv M^{\alpha\beta}_{~~~~\mu\nu}D_{\beta}u^{\nu}\,,\quad
a^{\mu} \equiv u^{\alpha}D_{\alpha}u^{\mu}.
\end{equation}
From Eqs.(\ref{2.4b}) and (\ref{2.4c}),  we find that
\bq
\lb{2.7}
\lambda = u_{\beta}D_{\alpha}J^{\alpha\beta} + c_4 a^2 - 8\pi G  T_{\alpha}u^{\alpha},
\eq
where $a^{2}\equiv a_{\lambda}a^{\lambda}$.

Recently,  the combination of the gravitational wave event GW170817 \cite{GW170817}, observed by the LIGO/Virgo collaboration, and the event of the gamma-ray burst 
GRB 170817A \cite{GRB170817} provides  a remarkably stringent constraint on the speed of the spin-2 mode, $- 3\times 10^{-15} < c_T -1 < 7\times 10^{-16}$. In the Einstein-aether theory, 
the speed of the spin-2 graviton is given by $c_{T}^2 = 1/(1-c_{13})$ \cite{JM04}, so the  GW170817 and GRB 170817A events imply  
\bq
\lb{2.8a} 
\left |c_{13}\right| < 10^{-15}.
 \eq
Together with other observational and theoretical constraints, recently it was found that the parameter space of the theory is further restricted to the ranges \cite{OMW18},
 \bq
\lb{2.8b} 
 0 \lesssim c_{14} \lesssim 2.5\times 10^{-5}, \quad c_{4} \lesssim 0, \quad 0 \lesssim c_2 \lesssim 0.095.
 \eq
It should be noted that not all the points inside these ranges satisfy all the observational and theoretical constraints, and additional conditions still  exist even inside these ranges. 
For example, for  $0\le c_{14}\leq 2\times 10^{-7}$ we must further require $c_{14} \lesssim c_2 \lesssim  0.095$; and for $2\times 10^{-6}\lesssim c_{14}\lesssim 2.5\times 10^{-5}$, we 
need to further require $0 \lesssim c_2-c_{14} \lesssim 2\times 10^{-7}$. 
For details, see \cite{OMW18}.

\subsection{Polarizations and Interaction of Gravitational Plane Waves}
  
 The spacetimes for gravitational plane waves can be cast in various forms, depending on the choice of the coordinates and gauge-fixing  \cite{Wang91th,SKMHH,Griffiths16}. 
 In this paper, we shall adopt the form originally due to Baldwin, Jeffery, Rosen (BJR) \cite{BJ,Rosen}, which can be cast as \cite{Wang91aa,Wang91th}
 \bqn
 \lb{3.1}
 ds^2 &=& -2e^{-M} du dv + e^{-U}\Big[e^{V}\cosh W dy^2 - 2\sinh W dydz \nb\\
 && + e^{-V}\cosh W dz^2\Big],
 \eqn
where  $M, U, V$ and $W$ are functions of $u$ only,  which in general represents a gravitational plane wave propagating along the null hypersurfaces $u = $ constant. 
The corresponding spacetimes belong to Petrov Type N  \cite{Wang91th,SKMHH,Griffiths16} \footnote{By rescaling the null coordinate $u \rightarrow u' = \int{ e^{-M(u)} du}$, without loss of the generality, 
one can always set $M = 0$.}. Choosing a null tetrad defined as,
 \bqn
 \lb{3.2}
 && l^{\mu} \equiv  B\delta^{\mu}_{v}, \quad n^{\mu} \equiv A\delta^{\mu}_{u}, \quad m^{\mu} =  \zeta^2 \delta^{\mu}_2 +  \zeta^3 \delta^{\mu}_3,\nb\\
 &&   
 \bar{m}^{\mu} = \overline{\zeta^2} \delta^{\mu}_2  + \overline{\zeta^3}  \delta^{\mu}_3,
 \eqn
 where $A$ and $B$ must be chosen so that $M \equiv \ln(AB)$, and 
 \bqn
 \lb{3.3}
 \zeta^2 &\equiv& \frac{e^{(U-V)/2}}{\sqrt{2}}\left(\cosh \frac{W}{2} + i \sinh \frac{W}{2} \right), \nb\\
  \zeta^3 &\equiv& \frac{e^{(U+V)/2}}{\sqrt{2}}\left(\sinh \frac{W}{2} + i \cosh \frac{W}{2} \right),
 \eqn
we find that the Weyl tensor has only one independent component, represented by $\Psi_4$, and is given by \cite{Wang91th}, 
 \bqn
 \lb{3.4}
C^{\mu\nu\alpha\beta} &=&  4 \Big[\Psi_4 l^{[\mu}m^{\nu]} l^{[\alpha}m^{\beta]}   + \bar{\Psi}_4  l^{[\mu}\bar{m}^{\nu]} l^{[\alpha}\bar{m}^{\beta]}\Big],\nb\\
\Psi_4 &=& -\frac{1}{2}A^2\Bigg\{\cosh W V_{uu} + \cosh W \big(M_u - U_u\big)V_u \nb\\
&& + 2\sinh W V_u W_u  + i \Big[W_{uu} + \big(M_u - U_u\big)W_u \nb\\
&& - \sinh W \cosh W V_u^2\Big]\Bigg\},
 \eqn
where $[A, B] \equiv (AB - BA)/2$, and $V_u \equiv \partial V/\partial u$, etc. To see the physical meaning of $\Psi_4$, following \cite{Wang91aa,Wang91th}, let us first introduce the orthogonal spacelike unit vectors,
$E^{\mu}_{(a)}\; (a = 2, 3)$, in the $(y, z)$-plane via the relations,
 \bqn
 \lb{3.5}
E_{(2)}^{\mu} \equiv \frac{m^{\mu} + \bar{m}^{\mu}}{\sqrt{2}}, \quad
E_{(3)}^{\mu} \equiv \frac{m^{\mu} - \bar{m}^{\mu}}{i\sqrt{2}},
 \eqn
we find that the Weyl tensor can be written in the form, 
 \bqn
 \lb{3.6}
C^{\mu\nu\alpha\beta} &=& \frac{1}{2}\Big[e_{+}^{\mu\nu\alpha\beta} \left(\Psi_4 + \bar{\Psi}_4\right) 
+ i e_{\times}^{\mu\nu\alpha\beta} \left(\Psi_4 - \bar{\Psi}_4\right)\Big],\nb\\
\eqn
where
 \bqn
 \lb{3.7}
 e_{+}^{\mu\nu\alpha\beta} &\equiv&  4\Big(l^{[\mu}E_{(2)}^{\nu]}l^{[\alpha}E_{(2)}^{\beta]} - l^{[\mu}E_{(3)}^{\nu]}l^{[\alpha}E_{(3)}^{\beta]}\Big), \nb\\
e_{\times}^{\mu\nu\alpha\beta} &\equiv&  4\Big(l^{[\mu}E_{(2)}^{\nu]}l^{[\alpha}E_{(3)}^{\beta]} + l^{[\mu}E_{(3)}^{\nu]}l^{[\alpha}E_{(2)}^{\beta]}\Big).
\eqn
 
Making a rotation in the $\left(E_{(2)}, E_{(3)}\right)$-plane,
\bqn
\lb{3.8}
E_{2} &=& E'_{(2)} \cos\varphi + E'_{(3)}\sin\varphi,\nb\\
E_{3} &=& - E'_{(2)} \sin\varphi + E'_{(3)}\cos\varphi,
\eqn
we find that
\bqn
\lb{3.9}
e_{+} &=& e'_{+} \cos2\varphi + e'_{\times}\sin2\varphi,\nb\\
e_{\times} &=& - e'_{+} \sin2\varphi + e'_{\times}\cos2\varphi.
\eqn
In particular, if we choose $\varphi$ such that 
\bqn
\lb{3.10}
\varphi = \frac{1}{2} \tan^{-1}\left(\frac{{\mbox{Im}}\left(\Psi_4\right)}{{\mbox{Re}}\left(\Psi_4\right)}\right),  
\eqn
we obtain 
 \bqn
 \lb{3.11}
C^{\mu\nu\alpha\beta}=  \frac{1}{2}\left|\Psi_4\right|  {e_{+}'}^{\mu\nu\alpha\beta}.
\eqn
Thus, the amplitude of the Weyl tensor is proportional to the absolute value of $\Psi_4$, and the angle defined by Eq.(\ref{3.10}) is the polarization angle of the gravitational plane wave in the 
plane spanned by $\left(E_{(2)}, E_{(3)}\right)$, which is orthogonal to the propagation direction $l^{\mu}$ of the gravitational plane wave. It is interesting to note that  
the unit vectors $E^{\mu}_{(2)}$ and $E^{\mu}_{(3)}$ are parallelly transported along $l^{\nu}$, 
 \bqn
 \lb{3.12}
l^{\nu} D_{\nu} E^{\mu}_{(2)} = 0  = l^{\nu} D_{\nu} E^{\mu}_{(3)}.
\eqn
Therefore, the  angle defined by Eq.(\ref{3.10})  is invariant with respect to the parallelly transported basis  $\left(E_{(2)}, E_{(3)}\right)$ along the propagation direction $l^{\mu}$ of the gravitational plane wave \footnote{Polarizations
of GWs in weak-field approximations were also studied in \cite{Gong18} in the framework of Einstein-aether theory.}.
This is an important property belonging only to single gravitational plane waves.

When $W = 0$, from Eq.(\ref{3.5}) we find that 
\bq
\lb{3.13}
{\mbox{Im}}\left(\Psi_4\right) = 0, \; (W = 0), 
\eq
and $\varphi = 0$. Then,  the polarization is along the $E^{\mu}_{(2)}$-direction, which is  usually referred to as the
``+" polarization, characterized by the non-vanishing of the function $V$. The other polarization of the gravitational plane wave, often referred to as the ``$\times$" polarization, is represented
by the non-vanishing of the function $W$, for which generically we have ${\mbox{Im}}\left(\Psi_4\right) \not = 0\; (W \not= 0)$ (cf. Fig. 1 given in \cite{Wang91aa}). 

When  $M, U, V$ and $W$ are functions of $v$ only,  the gravitational plane wave is now propagating along the null hypersurfaces $v = $ constant. In this case, by rescaling the null 
coordinate $v \rightarrow v' = \int{e^{-M(v)} dv}$,   one can always set $M(v) = 0$. 

When gravitational plane waves moving in both of the two null directions are present, the metric coefficients $M, U, V$ and $W$ are in general  functions of $u$ and $v$. An interesting case is the 
collision of two gravitational plane waves moving along the opposition directions, which generically produces spacetime
singularities due to their mutual focuses \cite{Wang91b}. Another remarkable feature is that one of the gravitational plane waves can serve as a medium for the other, due to their non-linear 
interaction, so the polarizations of the gravitational plane wave can be changed.
 The change of polarizations due to the nonlinear interaction is exactly a gravitational analogue
of the Faraday rotation, but with the other gravitational plane wave  as the magnetic field and medium   \cite{Wang91aa,Wang91th,Wang91th}.

\section{Linearly Polarized gravitational plane waves}
 \renewcommand{\theequation}{3.\arabic{equation}} \setcounter{equation}{0}

In this section, we shall consider  gravitational plane waves moving along the hypersurfaces $u = $constant only with one direction of polarizations, which are usually called linearly polarized gravitational plane waves. 
Without loss of the generality, we shall consider only gravitational plane waves with the ``+" polarization. Then, 
by rescaling the $u$ coordinate, without loss of the generality, we can always set $M = 0$, so the metric takes the form,
 \bqn
 \lb{3.14}
 ds^2 &=& -2 du dv + e^{-U(u)}\Big(e^{V(u)} dy^2  + e^{-V(u)} dz^2\Big). ~~~~~
 \eqn
We also assume that the aether moves only in the ($u, v$)-plane, so its four-velocity $u_{\mu}$ takes the general form, 
\bq
\lb{3.15}
u^{\mu} = \frac{1}{\sqrt{2}} (e^{-h}, e^{h}, 0, 0).
\eq
Since the spacetime is only of $u$ dependence, it is easy to see that  $h = h(u)$. Then,  the  non-vanishing components of the Einstein and aether
  tensors $G_{\mu\nu}$ and $T^{\ae}_{\mu\nu}$ and the aether vector $\AE_{\mu}$ are given, respectively,
  by Eqs.(\ref{A.1}) and (\ref{A.2}).   In the vacuum case, we have $T^{m}_{\mu\nu} = 0,\; T_{\mu} = 0$, and the Einstein-aether equations (\ref{2.4a}) reduce to
 \bq
 \lb{3.18}
G_{\mu\nu} =  T^{\ae}_{\mu\nu},
\eq
which yield five equations \footnote{It is interesting to note that in Einstein's theory the field equations $G_{\mu\nu} = 0$ yields only a single equation \cite{Wang91aa,Wang91th}, 
\bq
\lb{GReq}
2U_{uu} - U_u^2  = V_u^2,  
\eq
for the two unknown functions $U(u)$ and $V(u)$. In this sense, the problem is underdetermined in Einstein's theory.  Thus, for any given
gravitational wave  $V(u)$, we can always integrate the above equation   to find $U(u)$.},  given by Eqs.(\ref{A.4a})-(\ref{A.4e}). 
The aether equations $\AE_{\mu} =0$ yield the same equation as given by Eq.(\ref{A.4c}).

It is remarkable to note that there are five independent field equations for the three unknowns, 
$U, V$ and $h$. Therefore, in contrast to the situation of  GR, in which  there is only one independent field equation, given by Eq.(\ref{GReq}),  for two unknown functions $U$ and $V$,  
here in the framework of the Einstein-aether theory, we are facing an overdetermined problem, instead of  underdetermined, and clearly
only for particular cases the above equations allow solutions for $U, V$ and $h$.  

From the constraint (\ref{2.8a}) we can see that the current observations of GW170817 and GRB 170817A  practically requires  $c_{13} \simeq 0$.  In addition, for the spin-2 gravitons  
to move precisely with the speed of light, we also need to set $c_{13} = 0$.  However, in order for our results to be as much applicable as possible,   in the rest of this section we shall not impose this condition, 
and consider all the possible solutions with both $c_{13}   = 0$ and  $c_{13} \not= 0$, separately.  
 
 \subsection{Solutions with $c_{13} = 0$}

When  $c_{13} = 0$,  Eqs.(\ref{A.4a})-(\ref{A.4e}) reduce to,
 \bqn
 \lb{3.20a}
&&  2 U_{uu} -\Big(U_u^2 + V_u^2\Big) \nb\\
&& ~~~~ +  2c_{14}\Big(h_{uu}   - h_u U_u - h_u^2\Big) = 0, \\
\lb{3.20b}
 && c_2 \Big(U_{uu} - 2h_u U_u - U_u^2\Big)     \nb\\
 && ~~~~ + \big(c_2   - c_{14}\big)\Big(h_{uu} - h_u U_u  - 2h_u^2\Big) = 0, \\
\lb{3.20c}
&& c_2 U_{uu} +  \big(c_2  - c_{14}\big)\Big(h_{uu} - h_u U_u   - h_u^2\Big) =0, \\
\lb{3.20d}
 &&   c_2\Big(2U_{uu} - U_u^2 - 4h_uU_u\Big) \nb\\
 && ~~~~ +  2 c_2h_{uu}  - \big(3c_2 + c_{14}\big)h_u^2 = 0.
\eqn
Then, from Eqs.(\ref{3.20b}) and (\ref{3.20c}) we find 
 \bqn
 \lb{3.21a}
&&   c_2\Big(U_u^2 + 2U_uh_u\Big) + \big(c_2 - c_{14}\big) h_u^2 = 0,\\
\lb{3.21b}
&&   c_2\Big(U_{uu} + U_u^2\Big) + \big(c_2 + c_{14}\big) U_u h_u +  \big(c_2 - c_{14}\big) h_{uu} = 0.\nb\\
 \eqn
 
To study the above equations further, we need to distinguish the cases $c_2 \not=  c_{14}$ and  $c_2 = c_{14}$, separately.

\subsubsection{$c_2 \not=  c_{14}$}

In this case, from Eqs.(\ref{3.21a}) and (\ref{3.21b}) we find that
 \bqn
\lb{3.22a}
&&  h_u^2 = \frac{c_2}{c_{14} - c_2}\Big(U_u^2 + 2U_uh_u\Big),\\
\lb{3.22b}
&&   h_{uu} =  \frac{1}{c_{14} - c_2} \Bigg\{c_2\Big(U_{uu} + U_u^2\Big) + \big(c_2 + c_{14}\big) U_u h_u\Bigg\}.\nb\\
 \eqn
 Inserting the above expressions into Eq.(\ref{3.20d}), we find 
 \bq
\lb{3.23}
c_2c_{14} \Big(U_{uu} - U_u^2 - 2U_u h_u\Big) = 0,
 \eq
from which we can see that there are three different cases that need to be considered separately, 
 \bq
\lb{3.24}
i)\; c_2c_{14} \not= 0, \quad ii)\; c_2 = 0, \; c_{14} \not= 0, \quad   iii)\; c_2 \not= 0, \; c_{14} = 0. 
 \eq

{\bf Case i) $ c_2c_{14} \not= 0$:} In this case we have 
\bq
\lb{3.24a}
U_{uu} = U_u^2 + 2U_u h_u, 
\eq
which has the solution 
 \bq
\lb{3.25}
U_u = \alpha_0 e^{U+ 2h},   
 \eq
 where $\alpha_0$ is an integration constant. Then Eq.(\ref{3.20b}) reduces to
 \bq
 \lb{3.25a}
 h_{uu}-2h_u^2-h_uU_u=0,
 \eq
 which has the solution
 \bq
 \lb{3.25b}
 h_u=\alpha_1 e^{2h+U},
 \eq
 where $\alpha_1$ is an integration constant. Notice that
$ h_u\propto U_u$.
 In fact we may write 
 \bq
 \lb{3.25d}
 h=\alpha U + h_0, 
 \eq  
 where $\alpha$ and $h_0$  are constants. By substituting Eqs.(\ref{3.24a}) and (\ref{3.25d}) into Eq.(\ref{3.20c}) or (\ref{3.20d}) we find that
 \bq
 \lb{3.25e}
  \alpha= -\frac{\sqrt{c_2}}{\sqrt{c_2}\pm\sqrt{c_{14}}}.
 \eq
 By substituting Eqs (\ref{3.24a}) and (\ref{3.25d}) into Eq.(\ref{3.20a}) we find
 \bq
 \lb{3.25g}
 V=\beta U + V_0,
 \eq
 where $V_0$ is another integration constant, and
 \bq
 \lb{3.25h}
 \beta\equiv \pm\sqrt{1+4\alpha+2c_{14}\alpha^{2}}.
 \eq
 Now combining Eqs.(\ref{3.25d}) and (\ref{3.25}) we find
 \bq
 \lb{3.25f}
 U_u={\hat{\alpha}_{0}} e^{(2\alpha+1)U},
 \eq
 where
$\hat{\alpha}_{0}\equiv \alpha_{0} e^{2h_{0}}$.
Thus, we obtain 
 \bq
 \lb{3.25j}
 U(u)=-\frac{1}{2\alpha+1}\ln\left[-\alpha_0 (2\alpha+1)(u-u_0)\right], 
 \eq
 where $u_0$ is a constant of integration. Once $U(u)$ is given the functions $h(u)$ and $V(u)$ can be read off from Eqs.(\ref{3.25d}) and (\ref{3.25g}), respectively, that is, 
  \bqn
 \lb{3.25ja}
 V(u)&=&-\frac{\beta}{2\alpha+1}\ln\left[-\alpha_0 (2\alpha+1)(u-u_0)\right] + V_0, \nb\\
  h(u)&=&-\frac{\alpha}{2\alpha+1}\ln\left[-\alpha_0 (2\alpha+1)(u-u_0)\right] + h_0,\nb\\
 \eqn
 where   $\beta$ is given by Eq.(\ref{3.25h}) in terms of $\alpha$ and $c_{14}$.

 {\bf Case ii) $ c_2 = 0,\; c_{14} \not= 0$:} In this case  from Eqs.(\ref{3.20b}) and (\ref{3.20c})  we find that $h_u = 0$, that is
 \bq
\lb{3.27}
h(u) = h_0,
 \eq 
where $h_0$ is a constant. Then, Eqs.(\ref{3.20b}) - (\ref{3.20c}) are satisfied identically, while Eq.(\ref{3.20a}) reduce to
 \bq
\lb{3.27b}
2U_{uu} - U_u^2 = V_u^2, 
 \eq
which is the same as in GR, that is, in the present case the functions $U$ and $V$ are not uniquely determined. For any given $U(u)$, one can integrate the above equation to obtain $V(u)$. 

 {\bf Case iii) $ c_2 \not= 0,\; c_{14} = 0$:} In this case  from Eqs.(\ref{3.20b}) and (\ref{3.20c}) we find that  
 $U_u + h_u = 0$,  
which has the solution, 
 \bq
\lb{3.29}
 U = - h + U_0,
 \eq 
where $U_0$ is a constant. Inserting the above expression into Eq.(\ref{3.20b})  we find that $h_u = 0$, that is, 
 \bq
\lb{3.30}
 h = h_0.
 \eq 
Then, from Eq.(\ref{3.20a}) we obtain
  \bq
\lb{3.31}
 V = V_0,
 \eq
 where $V_0$ is a constant. By rescaling $y$ and $z$ coordinates, without loss of the generality, we can always set $V_0 = U_0 = 0$, so the solution represents the Minkowski spacetime.  That is, in the current
 case only the trivial Minkowski solution is allowed. 
 
\subsubsection{$c_2 = c_{14}$}

In this case, from Eq.(\ref{3.20c}) we find that
  \bq
\lb{3.32}
 c_2 U_{uu} = 0.  
 \eq
Therefore, depending on the values of $c_2$, we have two different cases.

{\bf Case i)  $c_2 =  c_{14} \not= 0$:} In this case, we must have $U_{uu} = 0$, which has the general solution,
 \bq
\lb{3.32a}
U(u) = \alpha_0  u + U_0,  
\eq
where $\alpha_0$ and  $U_0$ are two integration constants.  On the other hand, from Eq.(\ref{3.20b}) we find that 
 \bq
\lb{3.33}
h(u) = - \frac{\alpha_0}{2} u + h_0,  
\eq
while  Eq.(\ref{3.20d}) is satisfied identically.  Then, from Eq.(\ref{3.20a}) we find that
 \bq
\lb{3.33a}
V(u) = \pm \sqrt{\frac{(c_2 - 2)\alpha_0^2}{2}} \; u + V_0,   
\eq
 where $V_0$ is another integration constant.
 
{\bf Case ii)  $c_2 = c_{14} = 0$:} In this case, Eqs.(\ref{3.20b}) - (\ref{3.20d}) are satisfied identically for any given $h(u)$, while Eq.(\ref{3.20a}) reduces to
 \bq
\lb{3.34}
2U_{uu} - U_u^2 = V_u^2, 
 \eq
which is the same as in GR, that is, in the present case the functions $U$, $V$  and $h(u)$ are not uniquely determined. For any given $U(u)$ and $h(u)$, one can integrate Eq.(\ref{3.34}) to obtain $V(u)$.

 \subsection{Solutions with $c_{13} \not= 0$}

When $c_{13} \not= 0$, from Eqs.(\ref{A.4d}) and (\ref{A.4e}) we find that
\bq
\lb{3.35}
V_{uu} - U_uV_u - 2h_uV_u = 0,
\eq
which has the solution,
\bq
\lb{3.36}
V_u  = \alpha_0 e^{U + 2h}, 
\eq
where $\alpha_0$ is an integration constant. Inserting the above expression into Eqs.(\ref{A.4a})-(\ref{A.4e}), we obtain the following four independent equations for $U$ and $h$,
 \bqn
 \lb{3.37a}
&&  2U_{uu}-U_u^2 + 2c_{14}\Big(h_{uu} - h_u U_u - h_u^2\Big) = V_u^2, ~~~~~~~\\
\lb{3.37b}
 && c_2 \Big(U_{uu} - 2h_u U_u - U_u^2\Big)   \nb\\
&& ~~  + \big(c_2 + c_{13} - c_{14}\big)\Big(h_{uu} - h_u U_u - 2h_u^2\Big) = 0, ~~~~~ \\
\lb{3.37c}
&& 2\big(c_2 + c_{13} - c_{14}\big)\Big(h_{uu} - h_u U_u-h_u^2\Big)\nb\\
&& ~~~~~~~~~   +2c_2 U_{uu} + c_{13}U_u^2    += -c_{13}V_u^2, ~~~~~ \\
\lb{3.37d}
 &&   \big(c_{13} + 2c_2\big)\Big(2U_{uu} - U_u^2 - 4h_uU_u\Big)+  4 c_2h_{uu} \nb\\
 && ~~~~~~~~~ - 2\big(3c_2 - c_{13} + c_{14}\big)h_u^2 = - c_{13} V_u^2, ~~~~~
\eqn
Combining  Eqs.(\ref{3.37a})  and (\ref{3.37c}) we find
 \bq
 \lb{3.37e}
 c_{123}U_{uu}=(c_{13}c_{14}+c_2+c_{13}-c_{14})(h_u^2+h_u U_u-h_{uu}),
 \eq
 and by using Eqs.(\ref{3.37a})  and (\ref{3.37d}) we obtain 
 \bqn
 \lb{3.37f}
 && c_{123}U_u^2=(c_{13}c_{14}+2c_{13}-2c_{14})(h_u^2+h_u U_u-h_{uu})\nb\\
 &&~~~~~~~~~~+(c_{13}-c_{14}-c_2)h_u^2-2c_{123}h_u U_u.
 \eqn
 To study the above equations further, we need to consider separately the cases $c_{123}=0$ and $c_{123}\neq 0$.
 
 \subsubsection{$c_{123}=0$}
 
 In this case, from Eqs.(\ref{3.37a}) and (\ref{3.37c}) we find
 \bq
 \lb{3.37g}
 c_{14}(c_{13}-1)(h_{uu}-h_u^2-h_u U_u)=0.
 \eq
 The possibility of $c_{13}=1$ is ruled out by observation \cite{OMW18}, as mentioned above, leaving the possibilities
 \bq
 \lb{3.37h}
 c_{14}=0,
 \eq
 or
 \bq
 \lb{3.37i}
 h_{uu}-h_{u}U_{u}-h_u^2=0.
 \eq

{\bf Case A.1}   $c_{14}=0$: 
 In the case of Eq.(\ref{3.37h}) we find that Eqs.(\ref{3.37b}) and (\ref{3.37d}) reduce to
 \bq
 \lb{3.37j}
 U_{uu}=2h_u U_u+U_u^2,
 \eq
 and
  \bq
 \lb{3.37k}
 h_{uu}=2h_u^2+h_u U_u,
 \eq
 respectively, where we have used the fact that Eq.(\ref{3.37a}) reduces to $2U_{uu}=U_u^2+V_u^2$.  Then,  both $h_u$ and $U_u$ are proportional to $e^{2h+U}$, and hence by Eq.(\ref{3.36}) we find
\bqn
\lb{3.37l}
h=\alpha V + h_0 \quad
U=\beta V + U_0,
\eqn
where $h_0$ and $U_0$ are two integration constants, and the constants $\alpha$ and $\beta$ can be determined  by substituting Eq.(\ref{3.37l}) and Eq.(\ref{3.37j}) into Eq.(\ref{3.37a}) or Eq.(\ref{3.37c}), which yields   
\bq
\lb{3.37n}
\alpha = \frac{1- \beta^2}{4\beta}.
\eq
Inserting the above expressions into Eq.(\ref{3.36}), we find that
\bq
\lb{3.37na}
V = - \frac{2\beta}{1+ \beta^2} \ln\left[\hat\alpha_0\left(u_0 - u\right)\right],  
\eq
where $\hat\alpha_0 \equiv \alpha_0\left(2\alpha + \beta\right)e^{U_0+2h_0}$ and $u_0$ is an integration constant.
Therefore, in this case the solutions are given by Eqs.(\ref{3.37l})-(\ref{3.37na}).

{\bf Case A. 2}  $c_{14}\neq 0$:  In this  case  we find that
\bq
\lb{3.37o}
h_u=\alpha_1 e^{h+U},
\eq
and by Eq.(\ref{3.37d}) that
\bq
\lb{3.37p}
h_u^2\left(\frac{c_{14}}{c_{13}}-2\right)=0.
\eq

If $h_u=0$ ($\alpha_1=0$) then by Eq.(\ref{3.37b}) we have
\bq
\lb{3.37q}
U_u=\alpha_2 e^U,
\eq
and using this result with Eq.(\ref{3.37a}) we have
\bq
\lb{3.37r}
U =\pm V + U_0.
\eq
Inserting the above expressions into  Eq.(\ref{3.36}), we find that
\bq
\lb{3.37ra}
V = \mp \ln\left[\mp \hat\alpha_0(u-u_0)\right], 
\eq
where $\hat\alpha_0 \equiv \alpha_0 e^{2h_0+U_0}$ and where the choice of upper or lower sign must hold for both Eqs(\ref{3.37r}) and Eq.(\ref{3.37ra}). 
Thus, in this case, the general solutions are given by
\bq
\lb{casea.2.1}
\left(U, V, h\right) = \left(\pm V + U_0, V, h_0\right),
\eq
where $V$ is given by Eq.(\ref{3.37ra}), and $U_0$ and $h_0$ are two integration constants.

However,  if $h_u\neq 0$ then Eq.(\ref{3.37b}) reduces to
\bq
\lb{3.38*}
U_{uu}-2h_u^2-U_u^2-2h_u U_u=0,
\eq
and we add the LHS of Eq.(\ref{3.37i}) (which is zero) twice to the LHS of Eq.(\ref{3.38*}) to get
\bq
\lb{3.38a}
U_{uu}+2h_{uu}-4h_u^2-4h_u U_u-U_u^2=0,
\eq
which simplifies to
\bq
\lb{3.38b*}
2h_{uu}+U_{uu}=(2h_u+U_u)^2.
\eq
If we define a function f(u) such that
\bq
\lb{3.3c}
f(u)=2h(u)+U(u),
\eq
then Eq.(\ref{3.38b*}) can be written as
\bq
\lb{3.3d}
f_{uu}=f_u^2,
\eq
which has the solution
\bq
\lb{3.38e}
f=-\ln\left(-\alpha_3(u-u_0)\right),
\eq
where $\alpha_3$ and $u_0$ are integration constants.  If we multiply both sides of Eq.(\ref{3.37o}) by $e^h$ we have
\bq
\lb{3.38f}
h_u e^h=\alpha_1 e^{2h+U},
\eq
and making use of Eq.(\ref{3.38e}) we find
\bq
\lb{3.38g}
h_u e^h=-\frac{\alpha_1}{\alpha_3}\frac{1}{u-u_0},
\eq
whereupon integration we find
\bq
\lb{3.38h*}
h=\ln\left(-\frac{\alpha_1}{\alpha_3}\ln\left(u-u_0\right) +h_0\right).
\eq
So,  for the functions $U$ and $V$  we have
\bqn
\lb{3.38i}
U&=&-\ln(-\alpha_3(u-u_0))-2h,\\
\lb{3.38j}
V&=& -\frac{\alpha_0}{\alpha_3}\ln(u-u_0)+V_0.
\eqn
By substituting these results into Eq.(\ref{3.37a}) we find that $\alpha_3=\pm\alpha_0$.

\subsubsection{$c_{123}\neq 0$}

In this case we can substitute Eqs.(\ref{3.37e}) and (\ref{3.37f}) into Eq.(\ref{3.37b}) and by defining
\bq
\lb{3.38}
Q \equiv c_{123}-c_{14}+\frac{c_2}{c_{123}}(c_{13}-c_{14}-c_2),
\eq
we have
\bq
\lb{3.38a}
Q(h_{uu}-2h_u^2-h_u U_u)=0.
\eq
And so we must consider the cases where $Q\neq 0$ and $Q=0$.

{\bf Case B.1}  $Q\neq0$: Then, we have
\bq
\lb{3.38b}
h_u=\alpha_1 e^{2h+U}\propto V_u.
\eq
Using this result with Eqs.(\ref{3.37e}) and (\ref{3.37f}) we find also that
\bq
\lb{3.38c}
U_u=\alpha_2 e^{2h+U}\propto V_u,
\eq
and thus we can set
\bq
\lb{3.38d}
h=\alpha V + h_0, \quad 
U=\beta V + U_0,
\eq
for some constants $\alpha, \beta, h_0$ and $U_0$.  Substituting Eqs.(\ref{3.38d}) and (\ref{3.38e}) into Eqs.(\ref{3.37a}) and (\ref{3.37c}),  we find that   $\alpha$ and $\beta$ must satisfy the relations, 
\bqn
\lb{3.38f}
&& \beta^2 + 4\alpha\beta+2c_{14}\alpha^2 - 1 = 0, \\
\lb{3.38g}
&& 2(c_{14}-c_{13}-c_2)\alpha^2 -4c_2\alpha\beta \nb\\
&& ~~~~~ -(c_{13} +2c_2)\beta^2 - c_{13} = 0,
\eqn
which uniquely determine $\alpha$ and $\beta$, but the expressions for them are     too long to be presented  here. Inserting the above expressions into Eq.(\ref{3.36}), we find that
\bqn
\lb{3.38h}
V = - \frac{1}{2\alpha + \beta}\ln\left[\hat\beta_0(u_0-u)\right], 
\eqn
where $\hat\beta_0 \equiv \alpha_0(2\alpha + \beta) e^{U_0+2h_0}$. Therefore, in the present case, once $\alpha$ and $\beta$ are determined by Eqs.(\ref{3.38f}) and (\ref{3.38g}), 
the functions $V(u)$, $U(u)$ and the aether field $h(u)$ are given, respectively,  by Eqs.(\ref{3.38d}) and (\ref{3.38h}).

{\bf Case B.2}  $Q = 0$:
It will be helpful to try to solve for $c_{14}$ as a function of the other $c_i$'s, and to introduce a new parameter $\delta$ such that:
\bq
\lb{3.39h}
\delta=2c_2+c_{13}.
\eq
Then we find from Eq.(\ref{3.38}) that
\bq
\lb{3.39}
c_{14}\delta=c_{13}(c_{2}+\delta),
\eq
If we consider $\delta=0$, then we have $c_2=0$ since $c_{13}\neq 0$.  But by Eq.(\ref{3.39h}) this means we must have $c_{13}=0$, which violates our assumption, and so we \textit{must} have
\bq
\lb{3.39c}
\delta\neq 0,
\eq
and thus
\bq
\lb{3.39.1}
c_{14}=c_{13}\left(1+\frac{c_2}{\delta}\right),
\eq
is a general solution for the $Q=0$ case.   However, we can still have $c_2=0$ in general.  If that is the case then we have $c_{13}=c_{14}$ and we find from Eq.(\ref{3.37c}) that
\bq
\lb{3.39d}
V_u^2=-U_u^2,
\eq
and so to have real functions we must have U and V constant in u.  Then by considering Eqs.(\ref{3.37e}) and (\ref{3.37f}) with a vanishing $U_{u}$ we have
\bq
\lb{3.39e}
h_{uu}-h_u^2=0,
\eq
which has the solution
\bq
\lb{3.39f}
h=-\ln(\alpha(u-u_0)) + h_0,\; (c_2 = 0),
\eq
where $\alpha$ and $h_0$ are  the integration  constants.  So,  in the case of $c_2=0$ we have a static Minkowskian spacetime with a dynamical aether. 

 If $c_2\neq 0$,  then we find from Eqs.(\ref{3.37e}) and (\ref{3.37f}) that
\bq
\lb{3.39i}
U_{uu}-U_u^2=\frac{2c_2}{\delta}(h_u^2+h_u U_u-h_{uu})+\frac{2c_2}{\delta}h_u^2+2h_u U_u,
\eq
and
\bq
\lb{3.39j}
2U_{uu}-U_u^2=+\frac{2c_2}{\delta}h_u^2+2h_u U_u +  {\cal{D}}   (h_u^2+h_u U_u-h_{uu}),
\eq
where
\bq
\lb{d}
 {\cal{D}} \equiv  \frac{2c_2 c_{13}^2}{c_{123}\delta}+\frac{1}{\delta}(c_{13}^2+2c_2).
 \eq
 
These expressions can be substituted into Eqs.(\ref{3.37a}) and (\ref{3.37d}) to find
\bq
\lb{3.39k}
V_u^2=\left(c_{13}\frac{(c_2+\delta)}{c_{123}}-\frac{2c_2}{\delta}\right)+\frac{2c_2}{\delta}h_u^2+2h_u U_u,
\eq
and
\bq
\lb{3.39l}
V_u^2=\left(c_{13}\frac{(c_2+\delta)}{c_{123}}-\frac{2c_2}{c_{13}}\right)+\frac{2c_2}{\delta}h_u^2+2h_u U_u.
\eq
Equating these two gives us
\bq
\lb{3.39m}
c_2(h_{uu}-h_u^2-h_u U_u)=0.
\eq
Since now we have   $c_2\neq 0$,  then we must have
\bq
\lb{3.39n}
h_u=\alpha e^{h+U}.
\eq
In this case, Eq.(\ref{3.37c}) reduces to
\bq
\lb{3.39p}
V_u^2=-\frac{2c_2}{c_{13}}U_{uu}-U_u^2,
\eq
and by Eq.(\ref{3.37a}) we also have
\bq
\lb{3.39q}
V_u^2=2U_{uu}-U_u^2,
\eq
by the result of which we must have
\bq
\lb{3.39r}
U_{uu}=0,
\eq
since $c_{123}\neq 0$ in this case.  As $U_u$ must be a constant, then by Eq.(\ref{3.39q}) we find that $V_u$ must be also a constant, and to keep the constants real we must have $U_u$ and $V_u$ vanish, as before. 
Considering this result, Eq.(\ref{3.37b}) reduces to
\bq
\lb{3.39s}
h_u^2=0.
\eq
Therefore, when $c_2 \not= 0$, the spacetime must be Minkowski and the aether field is simply given by $h(u) = h_{0}$, this is, the solution in the present case is
\bq
\lb{3.39t}
\left(U, V, h\right) = \left(U_0, V_0, h_0\right),\; (c_2 \not= 0),
\eq
where $U_0, \; V_0$ and $h_0$ are all constants.

\section{Summary}
 \renewcommand{\theequation}{4.\arabic{equation}} \setcounter{equation}{0}

In this paper, we have studied  gravitational plane waves   in Einstein-aether theory, and found all vacuum solutions of the linearly polarized gravitational plane waves.
 In general, such waves need to satisfy five independent Einstein-aether field
equations, given by Eqs.(\ref{A.4a}) -(\ref{A.4e}), for three unknown functions $\left(U(u), V(u), h(u)\right)$. Therefore, the problem in the Einstein-aether theory is overdetermined, and it is expected
that gravitational plane waves exist only for some particular choices of the coupling constants $c_i$. This is sharply in contrast to Einstein's general relativity, in which the problem is actually 
underdetermined, i.e. the vacuum Einstein field equations $G_{\mu\nu}$ only yield one independent equation,
\bq
\lb{4.1}
2 U_{uu}  - U_u^2 = V_u^2,
\eq
for the two unknown functions $U$ and $V$. Thus, for any given $V(u)$, one can integrate Eq.(\ref{4.1}) to find the metric coefficient $U(u)$. This implies that  Einstein's theory allows the existence of 
any form of gravitational plane waves.  
This is no longer true in Einstein-aether theory, due to the presence of the time-like aether field. In particular,  in Einstein-aether theory in order to have arbitrary forms of gravitational plane waves
exist, the coupling constants $c_i$ must be chosen so that one of the following two conditions must be satisfied,  
\bqn
\lb{4.2}
&& (i) \;\; c_{13} = c_2 = 0, \;\; c_{14} \not= 0, \;\; h(u) = h_0,  \quad {\mbox{or}}\nb\\
&&  (ii)\;  c_{13} = c_2 = c_{14} = 0, \;\;\;\;\;\;\;\;  \forall \; h(u).
\eqn
In the former case it can be seen that the aether must be a constant, while in the latter the aether has no contributions to the spacetime, and $T^{\ae}_{\mu\nu} = 0$ identically, as can be seen from Eq.(\ref{A.1}). 

In addition to the above two cases, in which any form of gravitational plane waves are allowed to exist in Einstein-aether theory, there exist also several particular cases in which the spacetime and the aether field take particular
forms. In particular,  non-trivial solutions exist in the other six  particular cases,
\bqn
\lb{4.2}
 (iii) && \;\; c_{13} = 0, \; c_2\not= c_{14}, \; c_2 c_{14} \not= 0, \nb\\  
 (iv) &&  \;\; c_{13} = 0, \; c_2 = c_{14}  \not= 0, \nb\\  
 (v)  &&  \;\; c_{13} \not= 0, \; c_{123} =  c_{14}  = 0, \nb\\  
 (vi) && \;\; c_{13} \not= 0, \; c_{123} = 0, \;  c_{14}  \not= 0,h_u= 0, \nb\\  
 (vii) && \;\; c_{13} \not= 0, \; c_{123} = 0, \;  c_{14}  \not= 0,h_u\not= 0, \nb\\  
 (viii) && \;\;  c_{13} \not= 0, \; c_{123} \not= 0,  \; Q  \not= 0, 
\eqn
in which the particular solutions of the  vacuum  Einstein-aether field equations are given, respectively, by Eqs.(\ref{3.25j})-(\ref{3.25ja});  Eqs.(\ref{3.32a})-(\ref{3.33a});
Eqs.(\ref{3.37l})-(\ref{3.37na}); Eqs.(\ref{3.37ra})-(\ref{casea.2.1}); Eqs.(\ref{3.38h*})-(\ref{3.38j}), and Eqs.(\ref{3.38d})-(\ref{3.38h}), where $Q$ is defined by Eq.(\ref{3.38}).  

In  the rest of the cases, the solutions are either not allowed or simply represent the Minkowski spacetime with either a constant or dynamical aether field.

Some of these cases are problematic, as outlined in Jacobson's review article \cite{Jacobson}.  Any case in which $c_{123}=0$ results in $\alpha_2$ diverging 
(suggesting that the current PPN analysis is not valid here), while any case in which $c_{14}=0$ results in the speeds of the scalar and vector modes diverging 
(suggesting that wave equations for these modes do not exist).

In Case {(iv)}, the squared speed of the spin-0 mode is given by  $c_S^2 = (2-c_2)/(2+3c_2)$. Thus, to have $c_S \ge 1$, we must require $c_2 = c_{14} < 0$, which is in conflict with the observational 
constraints of Eq.(\ref{2.8b}). Therefore, this case is ruled out by observations.

 If we require that the speeds 
of the scalar, vector and tensor modes are all precisely  equal to one, then we find that
\bq
\lb{4.3}
c_{13} = c_4 = 0, \quad
c_2 = \frac{c_1}{1-2c_1}, \;\; (c_T = c_V = c_S = 1),
\eq
which is satisfied only by  Case {(iii)}, and the corresponding solutions   are still quite different from those of GR, even all of these gravitational  modes now move
at the same speed as that of  the spin-2 graviton in GR.

It should be noted that the results obtained in this paper is quite understandable, since the aether field is always unity and timelike, while the gravitational plane waves move only along a null direction. Then, 
due to their mutual scattering, it is expected that oppositely moving gravitational  plane waves exist generically, and the spacetimes must depend on both $u$ and $v$. Therefore, if only a single gravitational wave
moving along a fix null direction is allowed to exist, it is clear that only for particular choices of the coupling constants $c_i$'s, can compatible solutions exist. 
 
Thus, it would be very interesting to study the interactions of a plane gravitational wave with the aether and  other matter fields, as well as
with a gravitational plane wave moving in the opposite direction, by paying particular attention on Faraday rotations and  the difference from those found in   GR \cite{Wang91aa,Wang91th,Wang91th}, 
due to the presence of the timelike aether field, which violates LI.

\section*{\bf Acknowledgements}
 
We would like to thank G. Cleaver, Bao-Fei Li and Bahram Shakerin for valuable discussions. The work of A.W. was supported in part by 
  the National Natural Science Foundation of China (NNSFC), Grant Nos. 11375153 and 11675145.

\section*{Appendix A: The Einstein and aether tensors $G_{\mu\nu}$ and   $T^{\ae}_{\mu\nu}$ and the aether vector  $\AE_{\mu}$}
 \renewcommand{\theequation}{A.\arabic{equation}} \setcounter{equation}{0}

For the spacetime of Eq.(\ref{3.14}), the  non-vanishing components of the Einstein tensor $G_{\mu\nu}$ and $T^{\ae}_{\mu\nu}$ are given by,
 \bqn
 \lb{A.1}
 G_{00} &=& \frac{1}{2}\Big(2U_{uu} - U_u^2 - V_u^2\Big), \nb\\
T^{\ae}_{00} &=&-\frac{1}{8}\Big[2c_2 U_{uu} + c_{13}\Big(V_u^2 + U_u^2\Big) \nb\\
&& + 2\big(c_{13} + c_2 + 3c_{14}\big)\Big(h_{uu} - h_u U_u - h_u^2\Big)\Big],\nb\\
T^{\ae}_{01} &=&\frac{e^{-2h}}{4}\Big[c_2 \Big(U_{uu} - 2h_u U_u - U_u^2\Big)   \nb\\
&& + \big(c_2 + c_{13} - c_{14}\big)\Big(h_{uu} - h_u U_u - 2h_u^2\Big)\Big],\nb\\
T^{\ae}_{11} &=&-\frac{e^{-4h}}{8}\Big[2c_2 U_{uu} + c_{13}\Big(U_u^2 +V_u^2\Big)   \nb\\
&& + 2\big(c_2 + c_{13} - c_{14}\big)\Big(h_{uu} - h_u U_u - h_u^2\Big)\Big],\nb\\
T^{\ae}_{22} &=& \frac{e^{V-U-2h}}{8}\Big[c_{13}\Big(2V_{uu} - V_u^2 - 2 U_uV_u - 4 h_uV_u\Big)    \nb\\
&& - \big(c_{13} + 2c_2\big)\Big(2U_{uu} - U_u^2 - 4h_uU_u\Big)\nb\\
&& - 4 c_2h_{uu} + 2\big(3c_2 - c_{13} + c_{14}\big)h_u^2\Big],\nb\\
T^{\ae}_{33} &=& -  \frac{e^{-(V+U+2h)}}{8}\Big[c_{13}\Big(2V_{uu} + V_u^2 - 2 U_uV_u - 4 h_uV_u\Big)    \nb\\
&& + \big(c_{13} + 2c_2\big)\Big(2U_{uu} - U_u^2 - 4h_uU_u\Big)\nb\\
&& + 4 c_2h_{uu} - 2\big(3c_2 - c_{13} + c_{14}\big)h_u^2\Big],
\eqn
and $\AE_{\mu} = \big(\AE_{0}, \AE_{1}, 0, 0\big)$, where 
 \bqn
 \lb{A.2}
 \AE_{0} &=&  - \AE_{1} e^{2h} = - \frac{e^{-h}}{4\sqrt{2}}\Big[2c_2 U_{uu} + c_{13}\Big(U_u^2 +V_u^2\Big)   \nb\\
&& + 2\big(c_2 + c_{13} - c_{14}\big)\Big(h_{uu} - h_u U_u - h_u^2\Big)\Big].
\eqn

In the vacuum case, we have $T^{m}_{\mu\nu} = 0$, and the Einstein-aether equations (\ref{2.4a}) reduce to
 \bq
 \lb{A.3}
G_{\mu\nu} =  T^{\ae}_{\mu\nu},
\eq
which yield five independent equations,
 \bqn
 \lb{A.4a}
&&  2U_{uu} - \Big(V_u^2 + U_u^2\Big)+ 2c_{14}\Big(h_{uu} - h_u U_u - h_u^2\Big) = 0, ~~~~~~~\\
\lb{A.4b}
 && c_2 \Big(U_{uu} - 2h_u U_u - U_u^2\Big)   \nb\\
&& ~~~~  + \big(c_2 + c_{13} - c_{14}\big)\Big(h_{uu} - h_u U_u - 2h_u^2\Big) = 0, ~~~~~ \\
\lb{A.4c}
&& 2c_2 U_{uu} + c_{13}\Big(U_u^2 +V_u^2\Big)   \nb\\
&& ~~~~~  + 2\big(c_2 + c_{13} - c_{14}\big)\Big(h_{uu} - h_u U_u - h_u^2\Big) = 0, ~~~~~ \\
\lb{A.4d}
 && c_{13}\Big(2V_{uu} - V_u^2 - 2 U_uV_u - 4 h_uV_u\Big)    \nb\\
&& ~~~~~~~~  - \big(c_{13} + 2c_2\big)\Big(2U_{uu} - U_u^2 - 4h_uU_u\Big)\nb\\
&& ~~~~~~~~  - 4 c_2h_{uu} + 2\big(3c_2 - c_{13} + c_{14}\big)h_u^2 = 0, ~~~~~ \\
\lb{A.4e}
 && c_{13}\Big(2V_{uu} + V_u^2 - 2 U_uV_u - 4 h_uV_u\Big)    \nb\\
&& ~~~~~~~~  + \big(c_{13} + 2c_2\big)\Big(2U_{uu} - U_u^2 - 4h_uU_u\Big)\nb\\
&& ~~~~~~~~  + 4 c_2h_{uu} - 2\big(3c_2 - c_{13} + c_{14}\big)h_u^2 = 0 ~~~~
\eqn
where in Eq.(\ref{A.4a}) we have used the fact that $T^{\ae}_{00}$ can be expressed in terms of $T^{\ae}_{11}$ which is equal to zero.


\end{document}